\documentstyle[letter,mbf,epsfig,wrapft]{ptptex}

\def\bge{\begin{equation}}
\def\ene{\end{equation}}
\def\bg{\begin{eqnarray}}
\def\en{\end{eqnarray}}
\def\nn{\nonumber}

\title{Quark-meson coupling model with short-range quark-quark 
interactions}

\author{Koichi Saito and Kazuo Tsushima$^{*}$}

\inst{Tohoku College of Pharmacy, Sendai 981-8558 \\
$^*$Special Research Center for the Subatomic Structure of Matter \\
Department of Physics and Mathematical Physics \\
The University of Adelaide, Adelaide, SA 5005, Australia}

\recdate{
}

\abst{Short-range quark-quark correlations are introduced into the 
quark-meson coupling (QMC) model phenomenologically.  
We study the effect of the 
correlations on the structure of the nucleon in dense nuclear matter.
With the addition of correlations, the saturation curve
for symmetric nuclear matter is much improved at high density. 
}

\begin{document}

\maketitle

More than a decade ago, Guichon\cite{guichon} proposed a relativistic
quark model for nuclear matter, where it consists of non-overlapping
nucleon bags bound by the self-consistent exchange of scalar ($\sigma$)
and vector ($\omega$) mesons in mean-field approximation (MFA).
This model has been developed as the quark-meson
coupling (QMC) model, and successfully applied to various phenomena in
nuclear physics.\cite{qmc}  Recently, the model has been further 
extended using relativistic constituent quark models.\cite{const} 

So far, the use of the QMC model has been limited to the region of small to
moderate densities because it has been assumed that the
nucleon bags do not overlap.  It is therefore of great
interest to explore ways to extend the model to include
short-range quark-quark interactions, which
may occur when nucleon bags overlap at high density. 
In this report we will introduce these short-range correlations in
a very simple way, and calculate their effect on the quark 
structure of the nucleon in medium.  We refer to this model as 
the quark-meson coupling model with short-range correlations (QMCs).  

Let us consider uniformly distributed (iso-symmetric) 
nuclear matter with nuclear density $\rho_B$. 
At finite density the nucleon bags start to overlap with each other, and 
a quark in one nucleon may interact with quarks in other nucleons in the 
overlapping region.  
Since the interaction between the quarks is short range, it may be 
reasonable to treat it in terms of contact interactions. 
An additional interaction term of the form, 
${\cal L}_{int} \sim \sum_{i \neq j}{\bar \psi}_q(i) \Gamma_\alpha
\psi_q(i) {\bar \psi}_q(j) \Gamma^\alpha \psi_q(j)$, 
is then added to the original QMC lagrangian density.\cite{qmc}  
Here $\psi_q(i)$ is a quark field in the $i$-th nucleon and 
$\Gamma_\alpha$ stands for a combination of $\gamma$ matrices 
(with or without the isospin and color generators). 
In this report we shall consider only u and d quarks and simulate the 
short-range interaction using 
scalar- and vector-type couplings in MFA: 
$\Gamma_\alpha = 1$ and $\gamma_0$. 

Next we consider the probability for the nucleon bags to overlap, using a 
simple geometrical approach.\cite{geo}  
To measure the overlap of nucleons we treat them as the MIT bag with the 
radius $R$.  For any two nucleons the overlapping volume (measured in 
units of the nucleon volume $V_N(=4\pi R^3/3)$) is  
$V_{ov}(y) = 1 - 3y/4 + y^3/16$ (for $y \leq 2$),\cite{resc}  
and $V_{ov}$ vanishes beyond $y=2$, where $y=s/R$ with $s$ a 
distance between the two centers of the nucleon bags.  

In a nucleus with $A$ nucleons, any given nucleon may overlap with 
$(A-1)$ others.  If the nucleons are distributed according to some 
two-particle density function $\rho_2({\mbf r}_1, {\mbf r}_2)$, 
then the overlapping volume {\em per nucleon} (in units of $V_N$) 
is  
\bge
V_A = (A-1) \int \int d{\mbf r}_1 d{\mbf r}_2 
\rho_2({\mbf r}_1, {\mbf r}_2) V_{ov}(|{\mbf r}_1-{\mbf r}_2|/R), 
\label{va}
\ene
where the function $\rho_2$ may be given as the product of single-particle 
densities $\rho_1({\mbf r}_1)$ (normalized to unity), $\rho_1({\mbf r}_2)$ 
and the two-nucleon correlation function 
$F(|{\mbf r}_1-{\mbf r}_2|)$.  
In the limit $A \to \infty$, we then find 
the overlapping volume as 
$V_\infty = (\rho_B V_N) \times (1 - \xi)$ in the uniform matter, 
where 
\bge
\xi = \frac{1}{V_N} \int d{\mbf s} H(s) V_{ov}(s/R), \label{gzi}
\ene
with $H(s) = 1 - F(s)$.  Here $\xi$ describes the effect of two-nucleon 
correlation in the overlapping part.  
If there were no correlation between two 
nucleons (namely, $F(s)=1$ or $\xi=0$) $V_\infty$ were 0.32 (when 
$R=$ 0.8 fm and $\rho_0$ = 0.15 fm$^{-3}$, the saturation 
density of nuclear matter).  

For nucleons separated less than $\sim$1 fm, some modification of the 
two-nucleon density is expected and this is described by the correlation 
function $F(s)$.  In principle, one could calculate 
the correlation function within QMC self-consistently, as in the relativistic 
Brueckner-Bethe-Goldstone formalism etc.\cite{rbbg}  
It is, however, not easy.  In this exploratory study we shall use a 
phenomenological form of the correlation function.  

We use a convenient form of the correlation 
function proposed by Miller and Spencer\cite{miller} -- so-called the 
Miller-Spencer correlation function.  
We then find 
\bge
H(s) = 1 - F(s) = 1 - \left(1 - \frac{1}{4}g_p^2(y)\right) 
(1 + f(s))^2, \label{spencer}
\ene
where $g_p$ describes the Pauli (exchange) correlations and is given by 
$g_p(y) = (3/y) j_1(y)$, 
where $y = sk_F$ with the Fermi momentum $k_F$.  
The function $f(s)$ describes the 
strong repulsion in the core part and is parametrized as  
$f(s) = - (1 - \beta s^2) \exp(-\alpha s^2)$, 
where $\alpha$ and $\beta$ are parameters.\cite{miller}  

The Bethe-Goldstone theory places a restriction on the overall size of 
$f(s)$.  The nuclear matter density times integral of the square of $f(s)$ 
is known as the ``wound'' integral, $\kappa$: 
\bge
\kappa = \rho_B \int d{\mbf s}  f(s)^2.  \label{kap}
\ene
This quantity, which is the convergence parameter of the hole-line 
expansion, has a typical value of 0.12 around $\rho_B = \rho_0$.\cite{day} 
A next condition on $f(s)$ is obtained by the consistency 
requirement\cite{miller} 
\bge
\int d{\mbf s} \left(1 - \frac{1}{4}g_p^2(y)\right)
(2f(s)+f(s)^2) = 0, \label{cond1}
\ene
to ensure the normalization on $\rho_2$.  Another quantity of interest 
is the correlation length $\ell_c$ defined by\cite{miller} 
\bge
\ell_c = - \int_0^\infty ds (2f(s)+f(s)^2). \label{cond2}
\ene
We shall use the conditions (\ref{cond1}) and (\ref{cond2}) to 
determine the parameters $\alpha$ and $\beta$ and 
check the value of ``wound'' integral $\kappa$ 
numerically at the same time.  

Now it is natural to choose the probability of overlap  
to be proportional to $V_\infty$.  (We here ignore 
higher-configurations like 3-body correlation etc.) 
In MFA the Dirac equation for a quark field 
in a nucleon bag is then given by 
\bge
[i\gamma\cdot\partial - (m_q - g_\sigma^q\sigma - 
f_s^q \langle {\bar \psi}_q \psi_q \rangle) - (g_\omega^q\omega + 
f_v^q \langle \psi_q^\dagger \psi_q \rangle)\gamma_0] \psi_q = 0, 
\label{dirac}
\ene
where $m_q$ is the bare quark mass, 
$\sigma$ and $\omega$ are the mean-field values of the $\sigma$ and 
$\omega$ mesons and $g_\sigma^q$ and $g_\omega^q$ are, respectively,  
the $\sigma$- and $\omega$-quark coupling constants in the 
usual QMC model.\cite{qmc}  The new coupling constants 
$f_{s(v)}^q$ have been introduced
for the scalar (vector)-type short-range correlations, 
and are given by $f_{s(v)}^q = ({\bar f}_{s(v)}^q/M^2) V_\infty$ 
($M = 939$ MeV, the free nucleon mass). 
Note that since the coupling constants have dimension of (energy)$^{-2}$  
we introduce dimensionless coupling constants ${\bar f}_{s(v)}^q$. 
In Eq.(\ref{dirac}), $\langle {\bar \psi}_q \psi_q \rangle$ and 
$\langle \psi_q^\dagger \psi_q \rangle$ are, respectively, the average 
values of the quark scalar density and quark density with respect to 
the nuclear ground state, which 
are approximately given by the values at the center of the nucleon   
in local density approximation.\cite{qmc,blun} 

Now we can solve the Dirac equation Eq.(\ref{dirac}), as in the usual 
QMC, with the effective quark mass 
\bge
m_q^\star = m_q - (g_\sigma^q\sigma + 
f_s^q \langle {\bar \psi}_q \psi_q \rangle), 
\label{mstar}
\ene
instead of the bare quark mass.  The Lorentz vector interaction shifts the 
nucleon energy in the medium\cite{qmc} 
\bge
\epsilon({\mbf k}) = \sqrt{M^{\star 2} 
+ {\mbf k}^2} + 3(g_\omega^q \omega + f_v^q 
\langle \psi_q^\dagger \psi_q \rangle),  
\label{nenergy}
\ene
where $M^{\star}$ is the effective nucleon mass, which is given by 
the usual bag energy.\cite{qmc}  

The total energy per nucleon at density $\rho_B$ is then 
expressed as 
\bge
E_{tot} = \frac{4}{(2\pi)^3\rho_B} \int^{k_F} d{\mbf k}  
\sqrt{M^{\star 2}+ {\mbf k}^2} +  3(g_\omega^q \omega + f_v^q
\langle \psi_q^\dagger \psi_q \rangle) \nn \\ 
+ \frac{1}{2\rho_B} (m_\sigma^2 \sigma^2 - m_\omega^2 \omega^2), 
\label{etot}
\ene
where $m_\sigma$ and $m_\omega$ are 
respectively the $\sigma$ and $\omega$ meson masses.  
The $\omega$ field created by the uniformly distributed nucleons is 
determined by baryon number conservation: 
$\omega = 3g_\omega^q\rho_B / m_\omega^2 = 
g_\omega\rho_B / m_\omega^2$ (where $g_\omega = 3g_\omega^q$), while   
the $\sigma$ field is given by the thermodynamic condition: 
$(\partial E_{tot}/\partial \sigma) = 0$.  
This gives the self-consistency 
condition (SCC) for the $\sigma$ field\cite{qmc} 
\bge
\sigma = - \frac{4}{(2\pi)^3 m_\sigma^2} \left( \frac{\partial M^\star} 
{\partial \sigma} \right) \int^{k_F} d{\mbf k} \frac{M^\star}
{\sqrt{M^{\star 2}+ {\mbf k}^2}}, 
\label{scc}
\ene
where 
$(\partial M^\star / \partial \sigma) 
 = -3 g_\sigma^q S_N(\sigma) = - g_\sigma C_N(\sigma)$.  
Here $g_\sigma = 3g_\sigma^q S_N(0)$ and $C_N(\sigma) = 
S_N(\sigma)/S_N(0)$, with the quark scalar charge defined by $S_N(\sigma) = 
\int_N d{\mbf r} \ {\bar \psi}_q \psi_q$.  
In actual calculations, the quark density 
$\langle \psi_q^\dagger \psi_q \rangle$ in the total energy may be 
replaced by $3\rho_B$, and the quark 
scalar density, contributing to the effective quark mass, is 
approximately given as 
$\langle {\bar \psi}_q \psi_q \rangle = (m_\sigma^2/g_\sigma) \sigma$ 
because of the SCC.\cite{qmc,blun} 

\begin{figure}[t]
\parbox{\halftext}{
\epsfig{file=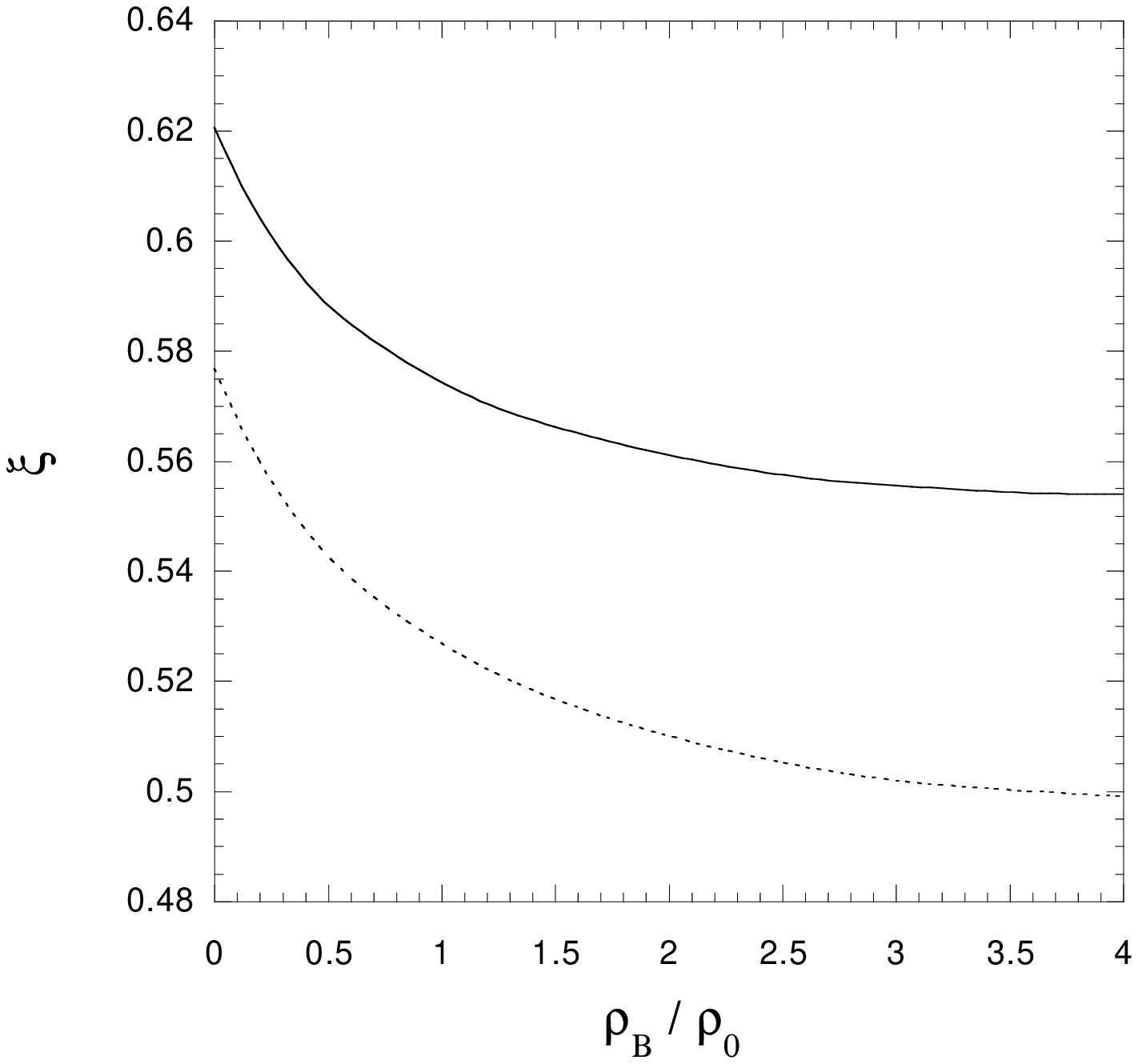,width=6.6cm}
\caption{$\xi$ as a function of $\rho_B$.  The solid (dotted) curve
is for $\ell_c$ = 0.75 (0.7) fm.}
\label{f:xi}}
\hspace{5mm}
\parbox{\halftext}{
\epsfig{file=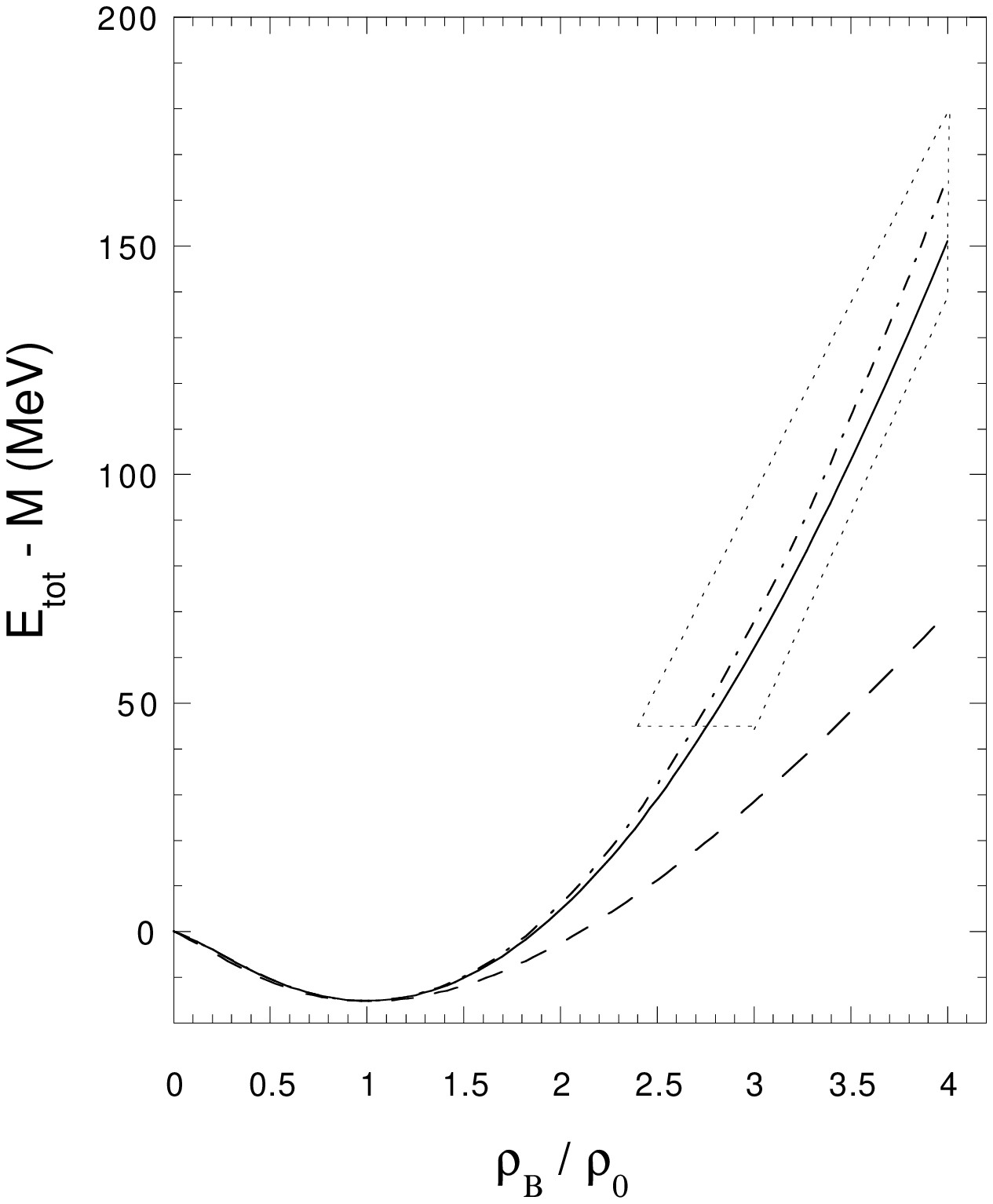,width=6.6cm}
\caption{Energy per nucleon for symmetric nuclear matter.
The dashed curve gives the result of the
original QMC.\protect\cite{qmc}  The solid and dot-dashed
curves are for $\ell_c$ = 0.75 and 0.7 fm in QMCs, respectively.
The region enclosed with the dotted curves
is the empirical one.\protect\cite{sano}}
\label{f:etot}}
\end{figure}
Now we are in a position to show our main results.  We choose $m_q$ = 5 MeV 
and the bag radius of the nucleon in free space $R_0$ to be 0.8 fm.  
The bag constant $B$ and usual parameter $z$, which accounts for the 
center of mass correction and gluon fluctuations, in the bag model are 
determined to fit the free nucleon mass with $R_0$ = 0.8 fm -- 
we find $B^{1/4}$ = 170.0 MeV and $z$ = 3.295.\cite{qmc}  
The coupling constants $g_\sigma$ and $g_\omega$ are determined so as 
to reproduce the average binding energy of symmetric nuclear matter 
($-15.7$ MeV) at the saturation density $\rho_0$.  

\begin{figure}[t] 
\parbox{\halftext}{
\epsfig{file=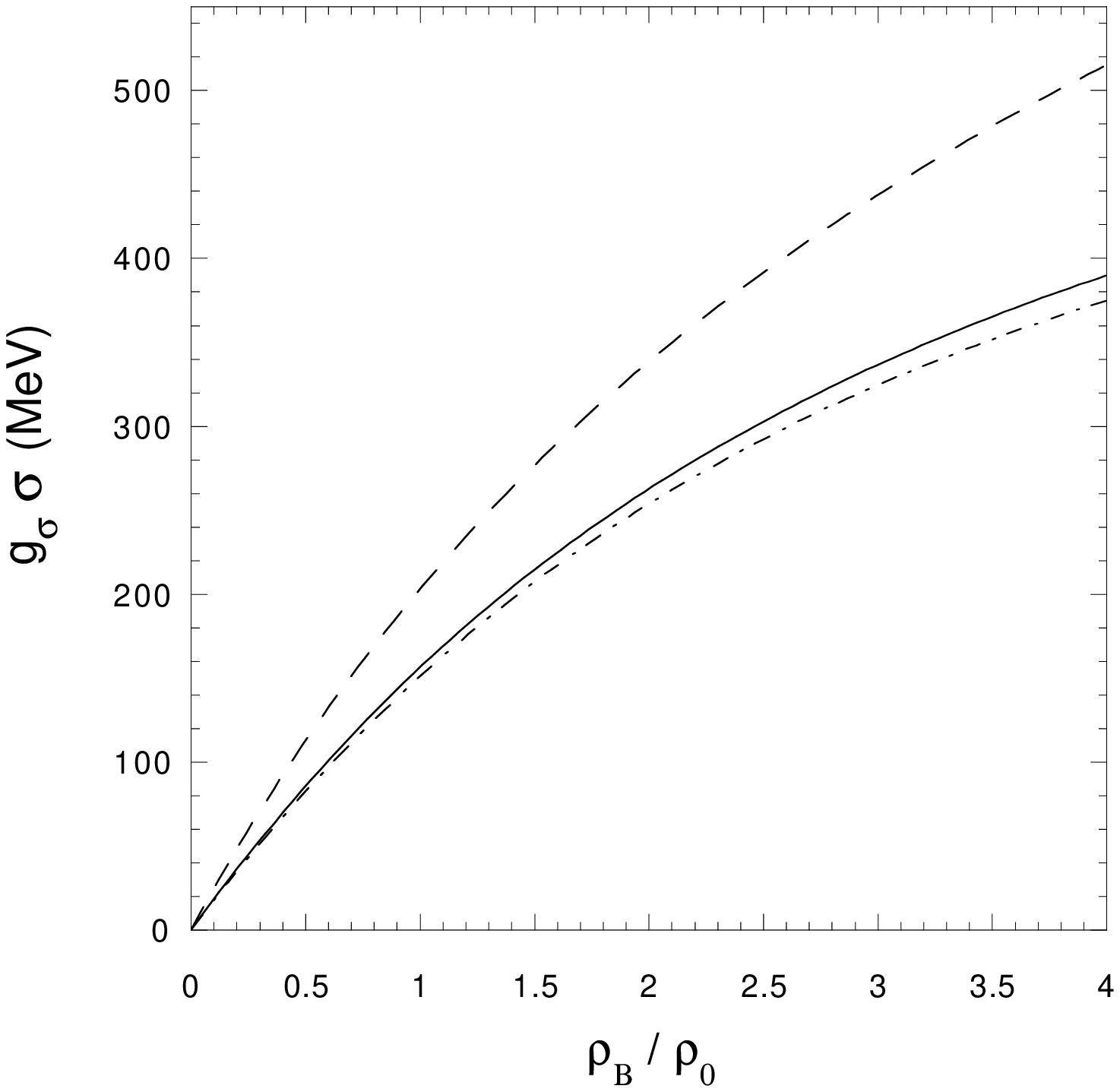,width=6.6cm}
\caption{Scalar mean-field values in nuclear medium.
The curves are labeled as in Fig.~\protect\ref{f:etot}.}
\label{f:gsig}}
\hspace{5mm}
\parbox{\halftext}{
\epsfig{file=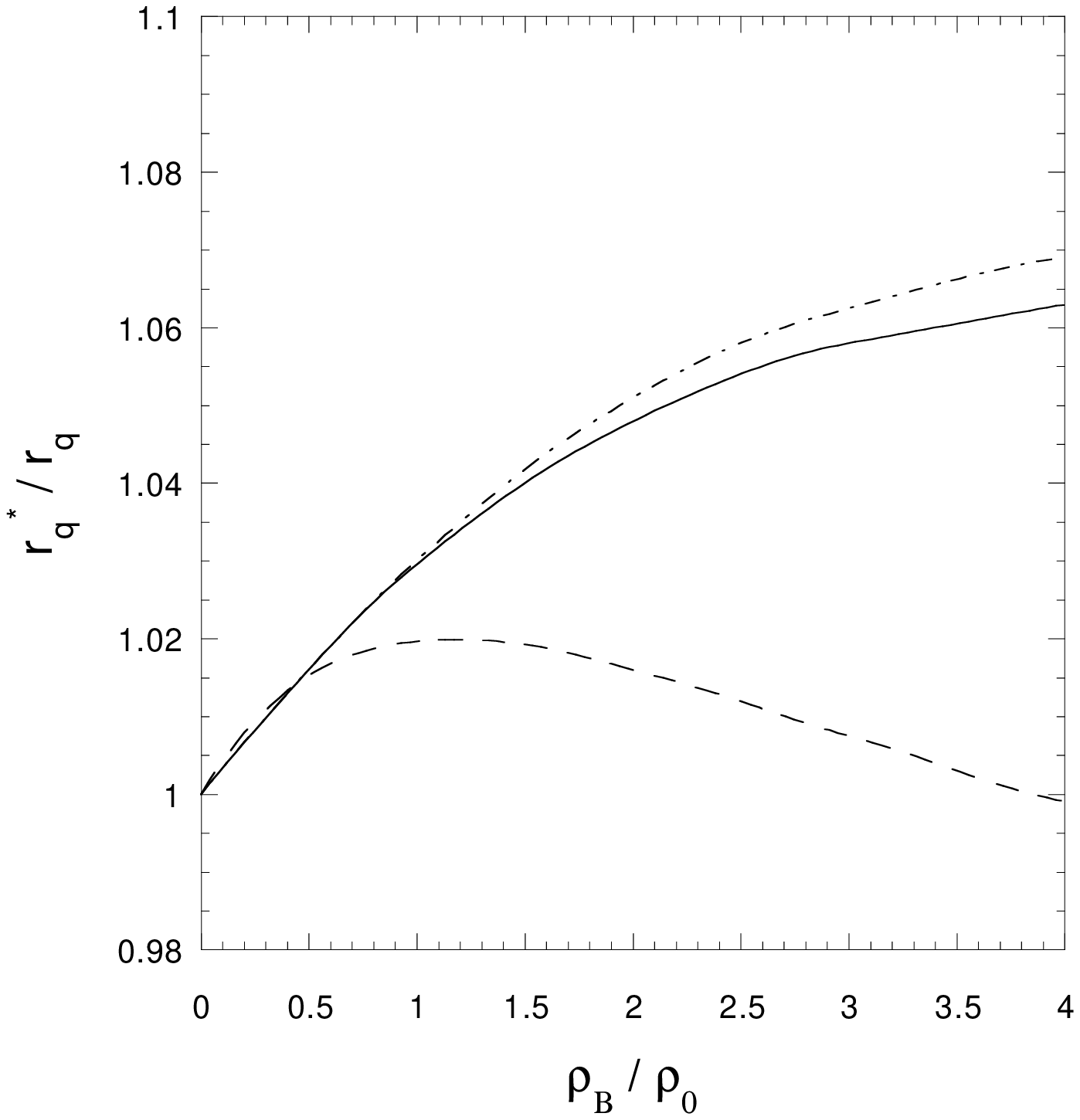,width=6.6cm}
\caption{Ratio of the in-medium rms radius of the nucleon $r_q^\star$ 
(calculated 
using the quark wave function) to that in free space ($r_q$ = 0.5824 fm).
The curves are labeled as in Fig.~\protect\ref{f:etot}.}
\label{f:rms}}
\end{figure}

In Fig.~\ref{f:xi}, we show $\xi$ in Eq.(\ref{gzi}).   
As an example, we set ${\bar f}_s^q = 40, {\bar f}_v^q = 8$
to fit the (observed) energy per nucleon at high $\rho_B$
(see Fig.~\ref{f:etot}).  If we fix 
$\ell_c$ = 0.75 (0.7) fm, the conditions (\ref{cond1}) and (\ref{cond2}) 
give $\alpha$ = 1.05 (1.20) fm$^{-2}$, $\beta$ = 0.617 (0.708) 
fm$^{-2}$ and $\kappa$ = 0.12 (0.10) at $\rho_0$.  
We then find that $g_\sigma^2$ = 50.09 (48.11), $g_\omega^2$ = 37.22 
(33.80) and $K=$ 360 (370) MeV.  
As shown in Fig.~\ref{f:xi}, $\xi$ gradually decreases 
and it becomes almost constant at high density.  

In Fig.~\ref{f:etot}, we present the saturation curve for symmetric 
nuclear matter.  We can see that QMCs can provide the energy per nucleon, 
which lies in the empirical region (enclosed with 
the dotted curves).\cite{sano}  
It may imply the importance of the short-range correlations at high density.
If we choose a large value of ${\bar f}_v^q$ the energy becomes
larger at high $\rho_B$ and it also makes the effective nucleon mass
large.  In QMCs we find $M^{\star}/M \sim 0.84$ at $\rho_0$. 
We also show the scalar mean-field values at finite 
density in Fig.~\ref{f:gsig}.  
In the original QMC the strength of the scalar field goes 
up to about 500 MeV at $\rho_B \sim 4 \rho_0$, while in QMCs it is much 
reduced at high $\rho_B$ because of the short-range (repulsive) interaction. 

Turning next to the size of the nucleon itself, as measured by 
the root-mean-square (rms) radius of the quark wave function, 
we see in Fig.~\ref{f:rms} that the short-range correlations give 
a little enhancement.  This effect is, however, not strong within 
the present parameter set, and the increase of the size around $\rho_0$ 
lies in the upper-limit value analysed by 
the electron scattering experiment.\cite{electron}  

We summarize the role of scalar- and vector-type 
short-range correlations.  The scalar-type 
correlation modifies the effective quark mass in a nuclear medium as 
(see Eq.(\ref{mstar})) 
\bge
m_q^{\star} = m_q - \left[ 1 + 
\frac{{\bar f}_s^q}{g_\sigma^2}\frac{m_\sigma^2}
{M^2}V_\infty(\rho_B) \right] (g_\sigma \sigma) . 
\label{efqmas}
\ene
In Eq.(\ref{efqmas}) the mean-field part 
($g_\sigma \sigma$ term) provides a density dependence on the 
quark mass of order of ${\cal O}(\rho_B)$ at low density, 
while the correlation part gives higher order contributions.  
This leads to the reduction of the scalar mean-field value in matter, 
which may affect the in-medium nucleon properties.\cite{qmc,qmc2} 

The vector-type correlation as well as the $\omega$ meson shifts 
the total energy.  The $\omega$ and the correlation contribute 
to the energy as (see Eq.(\ref{nenergy})) 
\bge
3(g_\omega^q \omega + f_v^q \langle \psi_q^\dagger \psi_q \rangle) 
= \frac{g_\omega^2}{m_\omega^2}  
\left[1+9\frac{{\bar f}_v^q}{g_\omega^2}\frac{m_\omega^2}
{M^2}V_\infty(\rho_B)\right] \rho_B.  \label{eshift}
\ene
Since $V_\infty$ depends on the density the correlation again gives 
the contribution of higher order in $\rho_B$.  It thus enhances 
the energy at high density.  If we introduce the effect of 
higher-configurations like 3-body, 4-body etc., such correlations may 
have dependences of higher power of $\rho_B$ and play an important role 
not only at high density but also near the region of the phase transition 
to quark-gluon plasma.  Those terms may correspond to higher order terms 
appearing in the chiral effective lagrangian for nuclear matter.\cite{gel}  

In conclusion, we have studied the effect of short-range quark-quark 
correlations associated with nucleon overlap in MFA.  
We have found that a repulsive vector-type correlation makes the saturation 
curve close to the empirical one at high density.  Furthermore, we have 
shown that the scalar- and vector-type correlations considerably modify 
the properties of nuclear matter at high density.  
While our inclusion of the correlations 
has been based on a quite simple, geometrical consideration, in the future 
we would hope to formulate the problem in a more sophisticated, dynamical 
way.\cite{rbbg}   

\section*{Acknowledgements}
The authors greatfully thank A.W. Thomas and G.A. Miller for valuable
discussions.
This work was supported by the Australian Research Council.

\end{document}